\documentclass[conference]{IEEEtran}
\IEEEoverridecommandlockouts
\usepackage{cite}
\usepackage{amsmath,amssymb,amsfonts}
\usepackage{algorithmic}
\usepackage{graphicx}
\usepackage{textcomp}
\usepackage{xcolor}
\usepackage{array}
\usepackage{tikz}
\usepackage{hyperref}

\def\BibTeX{{\rm B\kern-.05em{\sc i\kern-.025em b}\kern-.08em
    T\kern-.1667em\lower.7ex\hbox{E}\kern-.125emX}}

\newcommand\copyrighttext{%
  \footnotesize \textcopyright 2024 IEEE. Personal use of this material is permitted. Permission from IEEE must be obtained for all other uses, in any current or future media, including reprinting/republishing this material for advertising or promotional purposes, creating new collective works, for resale or redistribution to servers or lists, or reuse of any copyrighted component of this work in other works. 
  }
\newcommand\copyrightnotice{%
\begin{tikzpicture}[remember picture,overlay]
\node[anchor=south,yshift=10pt] at (current page.south) {\fbox{\parbox{\dimexpr\textwidth-\fboxsep-\fboxrule\relax}{\copyrighttext}}};
\end{tikzpicture}%
}

\begin{document}

\title{Advancing Airport Tower Command Recognition: Integrating Squeeze-and-Excitation and Broadcasted Residual Learning}

\author{\IEEEauthorblockN{Yuanxi Lin}
\IEEEauthorblockA{
\textit{Bauman Moscow State Technical University.}\\
Moscow, Russia \\
	linyu@student.bmstu.ru}
\and
\IEEEauthorblockN{Tonglin Zhou}
\IEEEauthorblockA{\textit{Wuhan University of Technology} \\
Hubei, China \\
}

\and
\IEEEauthorblockN{Yang Xiao}
\IEEEauthorblockA{
\textit{Fortemedia	Singapore}\\
Singapore\\
xiaoyang@fortemedia.com}}

\maketitle
\copyrightnotice

\begin{abstract}
Accurate recognition of aviation commands is vital for flight safety and efficiency, as pilots must follow air traffic control instructions precisely. This paper addresses challenges in speech command recognition, such as noisy environments and limited computational resources, by advancing keyword spotting technology. We create a dataset of standardized airport tower commands, including routine and emergency instructions. We enhance broadcasted residual learning with squeeze-and-excitation and time-frame frequency-wise squeeze-and-excitation techniques, resulting in our BC-SENet model. This model focuses on crucial information with fewer parameters. Our tests on five keyword spotting models, including BC-SENet, demonstrate superior accuracy and efficiency. These findings highlight the effectiveness of our model advancements in improving speech command recognition for aviation safety and efficiency in noisy, high-stakes environments. Additionally, BC-SENet shows comparable performance on the common Google Speech Command dataset.
\end{abstract}

\section{Introduction}
\label{sec:intro}

The airport tower is a crucial component of the air traffic control system, tasked with the real-time monitoring and direction of aircraft movements within the airspace. Tower speech commands~\cite{atc1,atc2} serve as an essential communication tool between tower commanders and pilots, playing a vital role in ensuring flight safety and optimizing airspace utilization. \\

Spoken keyword spotting~\cite{kws1,xiao2,xiao3} (KWS) is a task that can detect specific keywords within continuous speech. It has been widely used in voice assistants and other applications. Due to it always being powered on and always deploying into edge devices, it requires less power and have real-time latency than auto speech recognition models. Therefore it is ideal for tower speech command recognition because of its high accuracy and low latency. This method efficiently recognizes important commands from air traffic controllers, which are often standardized and repetitive. By identifying these key terms, keyword spotting ensures accurate and timely communication, enhancing flight safety and airspace efficiency. However, current keyword spotting systems face challenges, such as noisy environments and limited computational resources, especially in the busy settings of airport control towers. \\

Prior works on improving performance and noise robustness include using attention-based modules~\cite{convmixer,noisekws1} to enhance speech network efficiency. These modules help focus on valuable segments of the speech sequence. Additionally, self-attention methods, such as the keyword spotting transformer~\cite{kwt}, have outperformed convolutional network-attention hybrids. However, their high computational and memory demands limit their usability on small devices. Thus, there is a need for efficient KWS methods that are lighter but still effective. To address this, we explore lighter attention methods to improve KWS performance~\cite{senet,secrnn1}. We focus on enhancing the frequency and channel dimensions, which are crucial in the speech domain~\cite{secrnn2}. Therefore, we experiment with various attention methods on these dimensions to achieve better efficiency. \\

In this paper, we advance aviation command recognition using a deep-spoken keyword spotting model. The main contributions of this work are: 
\begin{itemize}
    \item We first create a dataset of standardized airport tower commands, including both routine and emergency instructions. By compiling this diverse dataset, we provide a robust foundation for training and evaluating spoken keyword spotting models. 
    \item We then enhance the broadcasted residual learning network~\cite{bcresnet,xiao1} with squeeze-and-excitation (SE) and time-frame frequency-wise squeeze-and-excitation (tfwSE) as efficient attention techniques, resulting in our BC-SENet model. This model focuses on crucial information with fewer parameters. 
    \item Testing on five keyword spotting models, including BC-SENet, demonstrates superior accuracy and efficiency. Our results show that BC-SENet improves command recognition in noisy environments, underscoring the value of our model for aviation safety and efficiency. Additionally, BC-SENet also performs well on the Google Speech Command dataset~\cite{gsc}. \\
\end{itemize} 
 
The remainder of the paper is organized as follows. Section \ref{sec:rel} reviews related works. Section \ref{sec:se} describes our BC-SENet model. Section \ref{ssec:ds} outlines the experimental setup, including configurations. Section \ref{sec:exp} details our dataset. Section \ref{resd} presents the results, comparing BC-SENet with other models. Section \ref{con} summarizes the key findings.

\section{Related work}
\label{sec:rel}

\textbf{Small Footprint Keyword Spotting -} Deep neural networks have proven effective in KWS tasks. Convolutional neural networks (CNNs) are popular in acoustic modeling because they can learn to encode spatial information from sequences. Earlier work~\cite{cnn} demonstrated CNNs' use for small footprint KWS. Recent efficient CNNs~\cite{tcresnet,matnet} often use repeated blocks based on residual learning and depthwise separable convolutions~\cite{dsres}. This trend continues in CNN-based KWS, using either 1D temporal or 2D frequency-temporal convolutions. Temporal convolutions require less computation but do not capture translation equivariance in the frequency dimension. Conversely, 2D convolutions still require more computation. Utilizing 1D temporal convolutions, TC-ResNet~\cite{tcresnet} requires fewer computations than 2D approaches. BC-ResNet~\cite{bcresnet} employs broadcast residual learning to address the inefficiency of 2D convolution and the limitations of 1D convolution. It applies frequency-wise 1D convolution to 2D audio features, averages these features over frequency to obtain temporal features, and then uses residual mapping where 1D residual information is broadcast. However, it is still limited to the ability to focus on crucial information in the frequency dimension.\\

\textbf{Attention mechanisms in CNNs} - The first gained attention with a visual attention method is proposed for image captioning. This led to numerous methods~\cite{van} focusing on attention mechanisms. A residual attention network introduced spatial attention using downsampling and upsampling~\cite{att}, while SENet~\cite{senet} proposed channel attention. SENet performs global average pooling on channels and calculates channel weights using fully connected layers. Inspired by these works, various studies~\cite{fmsg1,fmsg2} incorporated attention mechanisms to focus on valuable audio sequence segments. However, the high computational and memory demands limit their usability on small devices. Kim et. al.~\cite{secrnn1} explored alternative attention methods that are computationally efficient, considering channel and frequency dimensions to address the unique characteristics of the speech domain.

\section{Our BC-SENet Design}
\label{sec:se}

\begin{figure}[h]
    \centering
    \includegraphics[width=0.5\textwidth]{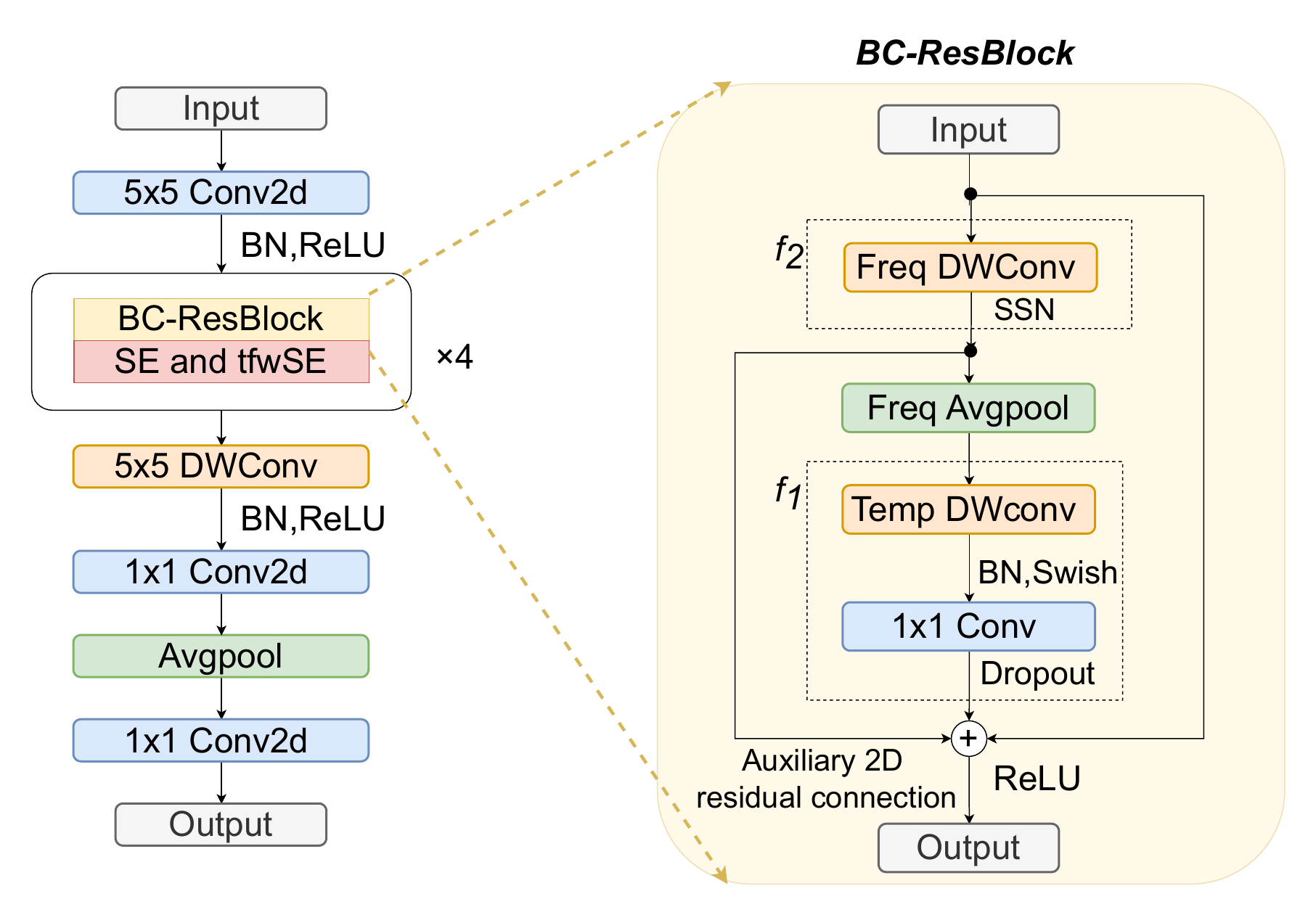} 
    \caption{Model architecture of the proposed BC-SEnet. The content in the yellow module describes the BC-ResBlock. The BC-ResBlock uses a frequency-depthwise convolution integrated with SSN. Then the feature undergoes averaging across the frequency dimension and is then processed through a temporal-depthwise separable convolution. Finally, the temporal feature is broadcasted back into 2D form at the residual connection.} 
    \label{fig:example} 
\end{figure}

\subsection{Network structure}
\label{ssec:subhead}

As shown in Figure 1, our entire architecture starts with a 5\(\times\)5 two-dimensional convolution layer, which extracts initial features through batch normalization and ReLU activation. This is followed by a series of Broadcasted Residual Blocks (BC-ResBlock)~\cite{bcresnet} that process complex features through depthwise separable convolutions while combining frequency and time domain operations to optimize feature transmission. Squeeze-and-excitation modules (SE and tfwSE) dynamically adjust the importance of channels to reweight features, enhancing the network's sensitivity to crucial information. Subsequent 5\(\times\)5 depthwise separable convolutions further process features, while 1\(\times\)1 convolutions are used for feature recombination and channel adjustment. Finally, average pooling reduces feature dimensions, and a 1\(\times\)1 convolution completes the output layer configuration to suit the final classification task. 

\subsection{BC-ResBlock}
The BC-ResBlock uses broadcasted residual learning to enhance feature extraction and efficiency in keyword spotting tasks by integrating temporal and frequency-domain processing. It starts with a frequency-depthwise convolution, processing input features along the frequency axis. SubSpectral Normalization (SSN)~\cite{ssn} then normalizes these features within subdivided frequency groups. The output is averaged across the frequency dimension and processed through a temporal-depthwise separable convolution. A residual connection adds the temporally processed output back to the original input through a broadcasting step, expanding the temporal features to the original input dimensions. This ensures alignment and combination of temporal and frequency features, enhancing the overall feature set. BC-ResBlock becomes
\begin{equation}
y = x + f_2(x) + BC(f_1(\text{avgpool}(f_2(x))))
\end{equation}
where \(x\) represents the input features to the residual block. \(f_2\) functions on 2D features, using depthwise separable convolution~\cite{dws} techniques, and processes data across both frequency and time dimensions to extract spatial characteristics of audio signals. \(f1\) operates on the temporal dimension, processes feature that have been simplified through \(f2\) and average pooling by frequency, converting 2D features into 1D features. \(BC\), which stands for \textbf{B}road\textbf{C}asting, refers to the operation that extends features to the frequency dimension, and \(avgpool\) represents the average pooling across the frequency dimension. \(y\) is the output of the residual block, which is the sum of the original input \(x\), the transformed features \(f_2(x)\), and the broadcasted features resulting from the processing pipeline \(f_1\) and average pooling applied to \(f_2(x)\). The ReLU activation function is applied after the residual. Following~\cite{bcresnet}, the channel of four BC-ResBlocks is [8, 12, 16, 20] and can be scaled to different layer widths. 
\subsection{SE and tfwSE block}
We use two lightweight attentions to enhance our model. First, the SE block consists of a squeeze operation followed by an excitation operation. The squeeze operation takes the average of the outputs from a 2D convolution across all dimensions except the channel dimension, creating a compressed intermediate representation. The excitation operation then employs two consecutive fully connected layers to generate attention weights, which indicate the significance of each channel relative to others.

\begin{figure*}[htb]
\begin{minipage}[b]{0.48\linewidth}
  \centering
  \centerline{\includegraphics[width=\linewidth]{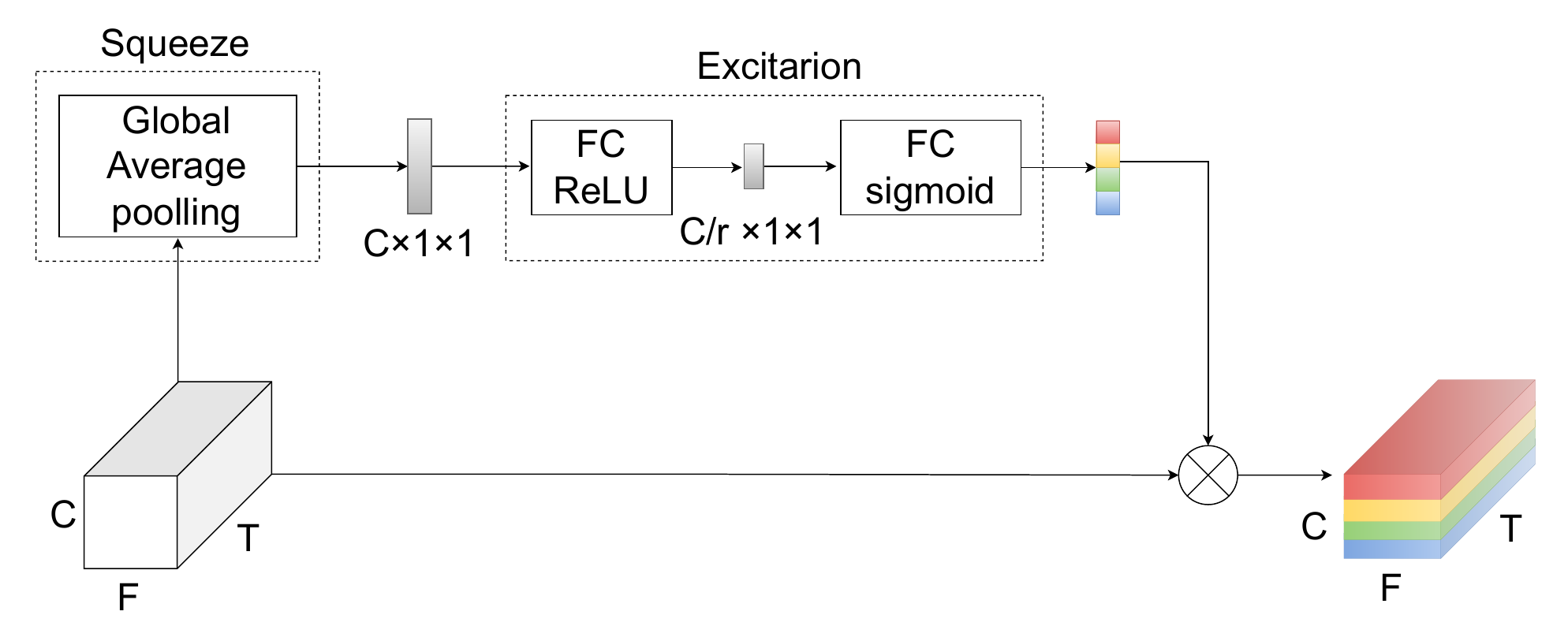}}
  \centerline{(a) Squeeze-Excitation}\medskip
\end{minipage}
\hfill
\begin{minipage}[b]{0.48\linewidth}
  \centering
  \centerline{\includegraphics[width=\linewidth]{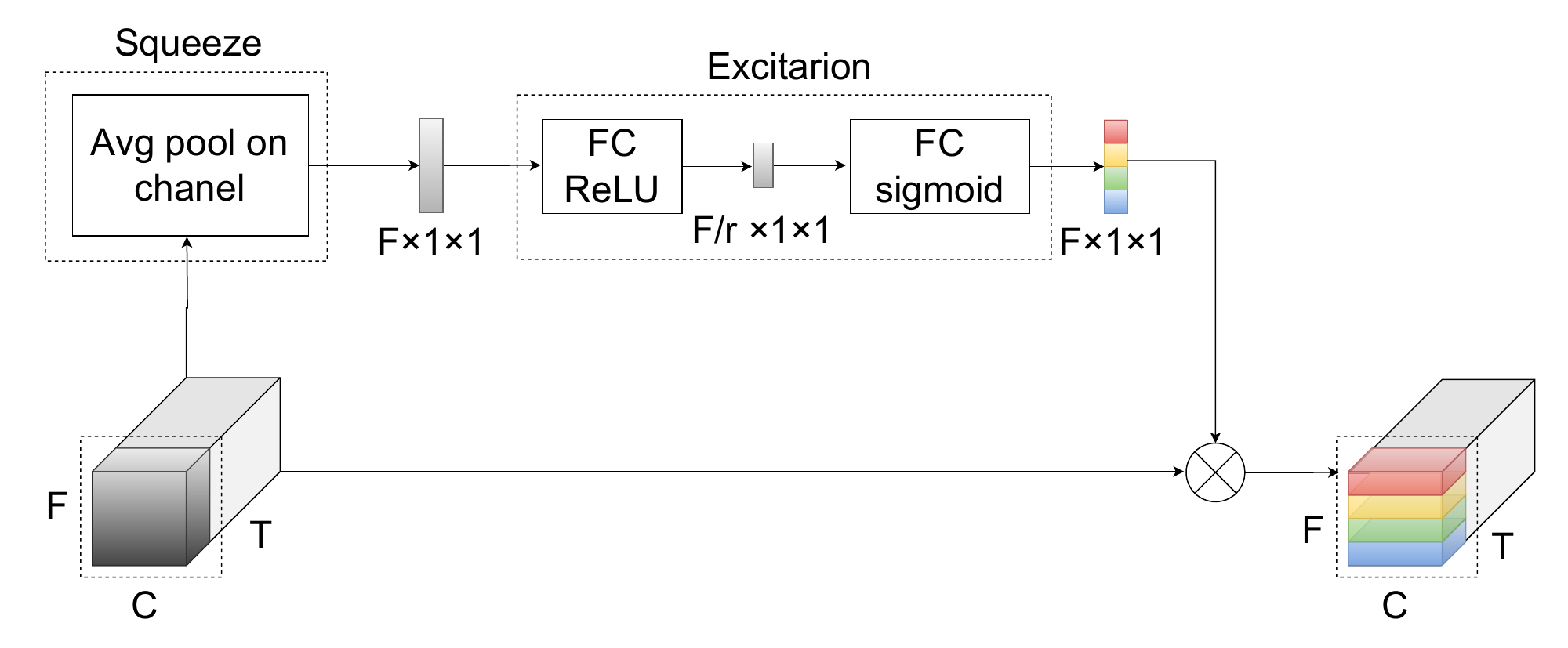}}
  \centerline{(b) Time-frame frequency-wise Squeeze-Excitation}\medskip
\end{minipage}
\caption{An illustration of SE block (a) and tfwSE block on one-time frame (b). The tfwSE applies this procedure for every time frame.}
\label{fig:res}
\end{figure*}

The basic SE mechanism, which focuses on reweighting channel importance via a two-step process of squeezing and excitation, is expanded to address unique characteristics of audio data, particularly its frequency content.
When applied to 2D audio data, squeeze operation is applied to the convolution output by
\begin{equation}
z_c = \frac{1}{F \times T} \sum_{f=1}^{F} \sum_{t=1}^{T} x_{cft}
\end{equation}

Where \(z_c\) is the squeezed intermediate representation for the 
\textit{c}-th channel. \(x_{cft}\) represents the output elements of a convolutional layer, \textit{c}, \textit{f}, and \textit{t} denote the channel, frequency, and time indices, respectively.\textit{F} and \textit{T} are the dimensions of frequency and time. This formula calculates the average of all frequency and time points' outputs for each channel, providing the input for the subsequent excitation operation.

The excitation operation is composed of two FC layers as follows:
\begin{equation}
y = \sigma(W_2 \delta(W_1 z))\
\end{equation}

Where \(y\) represents the output weights of the excitation operation, used to adjust the importance of each channel in the convolutional output. \textit{z} is the intermediate representation vector obtained from the squeeze operation. \(W_1\) and \(W_2\) are fully connected layers that learn and output the relative importance of each channel from the squeezed representation. \(\delta\) denotes the ReLU activation function, used to introduce non-linearity and enable the network to learn complex dependencies. \(\sigma\) is the Sigmoid function, ensuring that the output weights \(y\) are between 0 and 1, allowing these weights to directly adjust the contribution of each channel.

Second, the Time-frame Frequency-wise Squeeze-and-Excitation (tfwSE), is a specially designed attention mechanism used to process temporal and frequency information in audio signals. This method emphasizes applying attention weights independently on each time frame, rather than uniformly across the entire signal.

Within each time frame, tfwSE first calculates the average across channels for each frequency. This step simplifies the complexity of the input, producing a compressed representation of frequency that reflects the overall activity level at that specific time point. The squeeze operation on time frame t can be expressed by the following equation:
\begin{equation}
z_{ft} = \frac{1}{C} \sum_{c=1}^{C} x_{cft}
\end{equation}

Where \(z_{ft}\)is the squeezed intermediate representation for a specific frequency \textit{f} and time frame \textit{t}.

Next, a set of fully connected layers (typically two layers) is used to learn the important weights of frequencies from the squeezed representation. These weights are then used to adjust the strength of corresponding frequencies in the original convolutional output, thereby emphasizing or suppressing certain frequencies. Then excitation is applied on frequency dimension on each time frame as follows:

\begin{equation}
y_t = \sigma(W_2 \delta(W_1 z_t))
\end{equation}

Where \(y_t\) represents the attention weights output by the excitation operation for a specific time frame \(t\), used to adjust the importance of various frequencies in the convolutional output. \(y_t\) is the intermediate representation vector from the squeeze operation, specifically for time frame \(t\). This excitation operation processes the compressed vector for each time frame, generating weights that can adjust the frequency outputs of subsequent layers, thereby enhancing the model's focus on important features.

Our design combines BC-ResNet with Squeeze-and-Excitation methods, including tfwSE, to create a powerful framework that maximizes the strengths of both residual and attention-based learning. This framework integrates 1D and 2D convolutional features, effectively managing audio signals by capturing both temporal dynamics and frequency details. It achieves high accuracy without demanding extensive computational resources, making it suitable for devices with limited processing power. The attention mechanisms enhance accuracy and reduce errors by focusing on the most relevant features of sound signals and re-weighting input features to emphasize important information while downplaying less relevant data.

\section{Experiment setting}
\label{sec:exp}
\subsection{Datasets}
\label{ssec:ds}
\subsubsection{Chinese Tower Commands Dataset}
The collected tower control command keyword dataset consists of 17,783 audio files spanning 15 keywords. Each file is in WAV format, approximately 1.1 seconds long, and in 16-bit mono PCM format. The audio files have a 16KHz sampling rate, ensuring high-quality and precise audio data. The dataset includes voice recordings from 60 volunteers, all speaking Mandarin, with a male-to-female ratio of 7:3 and a balanced mix of southern and northern accents. The recordings were made at different times, distances from the microphone, and with varying tones and accents, adding diversity to the dataset. This variety helps in training models to better handle different voice information.

The \textit{Chinese Tower Commands} (CTC) dataset covers 12 common and 3 special tower control commands, providing a comprehensive set of tower control command characteristics. This makes it suitable for tower control command keyword recognition tasks. The detailed data volume of the 15 keywords in the dataset is shown in the table, with Chinese commands presented in Pinyin.
\begin{table}[tbp]
    \centering
    \caption{The table displays the number of audio files for each of the 15 tower control command keywords in the dataset. The commands are listed in Pinyin with their corresponding amounts.}
    \vspace{2mm}
    \resizebox{0.9\columnwidth}{!}{
    \begin{tabular}{|c|c|}
        \hline
        \textbf{Commands} & \textbf{Amount}  \\ \hline
        Kai Che & 1167 \\ \hline Hua Chu & 1190 \\ \hline Jin Pao Dao & 1189 \\ \hline
        Qi Fei & 1200 \\ \hline Zhuan Wan & 1152 \\ \hline Tiao Zheng Ju Li & 1190 \\ \hline
        Fu Fei & 1192 \\ \hline Shang Sheng Gao Du Tong Chang & 1184 \\ \hline Jie San & 1190 \\ \hline
        Fan Hang & 1191 \\ \hline Shou You Men & 1193 \\ \hline Ming Bai & 1188 \\ \hline
        Lun Chuan Bao Po & 1191 \\ \hline Xiao Su Du Li Lu & 1187 \\ \hline Sha Che Shi Xiao & 1179 \\ \hline
    \end{tabular}}
    \label{tab:commands}
\end{table}

\subsubsection{Google Speech Command Dataset}
We also conduct experiments on the \textit{Google Speech Command}
dataset v1 (GSC), which includes 64,727 one-second audio clips with 30 English keyword categories.  All of the audio clips in GSC are sampled at 16kHz in our experiment. There is the total of thirty words and we use ten classes of “Yes”, “No”, “Up”, “Down”, “Left”, “Right”, “On”, “Off”, “Stop”, and “Go” with two additional classes “Unknown Word (remaining twenty words)” and “Silence (no speech detected)” following the common settings.

\subsection{Implementation details}
We use input features of 40-dimensional log Mel spectrograms with a 30ms window size and a 10ms frameshift. The dropout rate is always set to 0.1. For the GSC dataset, all models are trained for 200 epochs using the stochastic gradient descent (SGD) optimizer with a momentum of 0.9, a weight decay of 0.001, a mini-batch size of 100, and a learning rate that increases linearly from zero to 0.1 over the first five epochs as a warmup, before decaying to zero with cosine annealing. For the CTC dataset, models are trained for 50 epochs using the Adam~\cite{adam} optimizer, a batch size of 64, with other settings remaining the same.

\begin{table}[tbp]
\centering
\caption{Performance comparison of various models on the \textit{Google Speech Commands} (GSC v1) and \textit{Chinese Tower Commands }(CTC) datasets, showing accuracy (\%) and parameter count for each model.}
\vspace{2mm}
\label{tab:clean}
\resizebox{0.9\columnwidth}{!}{%
\begin{tabular}{c|c|c|c}
\hline
\textbf{Model }             & \textbf{GSC v1} & \textbf{CTC}  & \textbf{Params} \\ \hline
BC-ResNet-1~\cite{bcresnet}        & 96.6   & 95.0 & 9.2K   \\
BC-ResNet-3~\cite{bcresnet}        & 97.6   & 98.0 & 54.2K  \\
BC-ResNet-6~\cite{bcresnet}        & 97.9   & 98.6 & 188K   \\
BC-ResNet-8~\cite{bcresnet}        & 98.0   & 98.7 & 321K   \\ \hline
DS-ResNet-14~\cite{dsres}       & 95.9   & 94.5 & 15.2K  \\
DS-ResNet-18~\cite{dsres}       & 96.7   & 98.3 & 72K    \\ \hline
TC-ResNet-8~\cite{tcresnet}        & 96.1   & 98.4 & 66K    \\
TC-ResNet-14~\cite{tcresnet}        & 96.2   & 98.6 & 137K   \\ \hline
MatchboxNet-3\(\times\)1\(\times\)64~\cite{matnet} & 97.2   & 98.3 & 77K    \\
MatchboxNet-3\(\times\)2\(\times\)64~\cite{matnet} & 97.5   & 98.4 & 93K    \\ \hline
\textbf{BC-SENet-1}        & 96.6   & 96.1 & 10K    \\
\textbf{BC-SENet-3}         & 97.7   & 98.4 & 61K    \\
\textbf{BC-SENet-6}         & 98.1   & 98.8 & 218K   \\
\textbf{BC-SENet-8}         & 98.2   & 99.1 & 376K   \\ \hline
\end{tabular}%
}
\end{table}

\section{Result and discussion}
\label{resd}
\subsection{Result on CTC dataset}
The CTC dataset results further highlight the robustness of our BC-SENet models refer to table \ref{tab:clean}. BC-SENet-8 achieves an impressive 99.1\% accuracy, the highest among all models tested, demonstrating its effectiveness in recognizing tower control commands. This model handles the sequential nature of the CTC dataset particularly well, which is crucial for real-world applications in speech recognition. BC-SENet-6 also performs exceptionally well with 98.8\% accuracy. In comparison, BC-ResNet models, while strong, do not reach the same accuracy levels, with BC-ResNet-8 at 98.7\%. DS-ResNet models show varied performance, with DS-ResNet-18 achieving 98.3\%. TC-ResNet models, especially TC-ResNet-14, perform strongly at 98.6\% but are still outperformed by BC-SENet models. MatchboxNet models, with MatchboxNet-3\(\times\)2\(\times\)64 reaching 98.4\%, also do not match the top-performing BC-SENet models. These results demonstrate that BC-SENet models offer superior accuracy and robustness, making them well-suited for tower control command recognition tasks in diverse and challenging environments.

\subsection{Result on GSC dataset}
The results for the GSC v1 dataset show that our BC-SENet models outperform the other models in terms of accuracy. BC-SENet-8 achieves the highest accuracy of 98.2\%, followed closely by BC-SENet-6 with 98.1\%. This demonstrates the effectiveness of integrating SE and tfwSE techniques. BC-ResNet models also perform well, with BC-ResNet-8 reaching 98.0\%, but they do not match the performance of BC-SENet models. DS-ResNet models, while having fewer parameters, achieve lower accuracy, with DS-ResNet-18 at 96.7\%. TC-ResNet models show decent performance, especially TC-ResNet-14 with 96.2\%, but again fall short compared to BC-SENet models. MatchboxNet models perform well, with MatchboxNet-3\(\times\)2\(\times\)64 achieving 97.5\%, but they do not surpass the BC-SENet models. Overall, the BC-SENet models demonstrate superior performance on the GSC v1 dataset, highlighting their ability to handle diverse and noisy data effectively.

\begin{table}[tbp]
\centering
\caption{Performance comparison of various models under different noise levels (-10dB, 0dB, 10dB), showing accuracy (\%) and parameter count for each model.}
\label{tab:noise}
\vspace{2mm}
\resizebox{0.9\columnwidth}{!}{%
\begin{tabular}{ccclc}
\hline
\textbf{Model}              & \textbf{-10dB} & \textbf{0dB}  & \textbf{10dB} & \textbf{Params} \\ \hline
BC-ResNet-1~\cite{bcresnet}        & 91.3  & 93.4 & 94.6 & 9.2K   \\
BC-ResNet-8~\cite{bcresnet}        & 97.5  & 98.6 & 98.7 & 321K   \\ \hline
DS-ResNet-18~\cite{dsres}       & 95.3  & 97.1 & 98.2 & 72K    \\ \hline
TC-ResNet-8~\cite{tcresnet}        & 96.4  & 97.5 & 98.1 & 66K    \\ \hline
MatchboxNet-3\(\times\)1\(\times\)64~\cite{matnet} & 95.3  & 97.3 & 98.0 & 77K    \\ \hline
\textbf{BC-SENet-1}         & 92.5  & 94.8 & 96.0 & 10K    \\
\textbf{BC-SENet-8}         & 98.1  & 98.7 & 98.7 & 376K   \\ \hline
\end{tabular}%
}
\end{table}

\subsection{Result on noisy CTC dataset}
The noise robustness analysis presents the performance of various models under different colored noise levels (-10dB, 0dB, and 10dB) of CTC dataset. BC-SENet-8 achieves the highest accuracy across all noise levels, demonstrating excellent noise robustness. At -10dB, BC-SENet-8 achieves 98.1\%, while at 0dB and 10dB, it maintains high accuracy at 98.7\%. This shows that BC-SENet-8 performs consistently well even in noisy environments. BC-ResNet-8 also performs strongly, with 97.5\% at -10dB and 98.6\% at 0dB. However, it does not surpass BC-SENet-8. DS-ResNet-18 and TC-ResNet-8 show good performance, with DS-ResNet-18 achieving 95.3\% at -10dB and 98.2\% at 10dB. TC-ResNet-8 achieves 96.4\% at -10dB and 98.1\% at 10dB. MatchboxNet-3\(\times\)1\(\times\)64 performs similarly, with 95.3\% at -10dB and 98.0\% at 10dB. BC-SENet-1, with fewer parameters, still shows good performance, achieving 92.5\% at -10dB and 96.0\% at 10dB. Overall, BC-SENet-8 demonstrates the best noise robustness, highlighting its effectiveness in maintaining high accuracy in noisy environments.
\section{Future direction}
Although the proposed BC-SENet demonstrates better performance on the Google Speech Commands v1 and our proposed Chinese Tower Commands datasets, it also requires more parameters because of the SE and tfwSE modules. Our further plan is to utilize the more parameter-efficient attention module. Efficient Channel Attention (ECA) is based on SE and aims to increase efficiency as well as decrease model complexity by removing the dimensionality reduction. We would like to integrate it with broadcast residual learning in future research.
\section{Conlusion}
\label{con}
This paper introduced BC-SENet for aviation command recognition. We created a dataset of standardized airport tower commands. BC-SENet combines Squeeze-and-Excitation and Temporal Frame-wise Squeeze-and-Excitation with broadcasted residual learning. Experiments showed BC-SENet's superior accuracy and efficiency on the Google Speech Commands v1 and our proposed Chinese Tower Commands datasets. BC-SENet also demonstrated excellent noise robustness. These results confirm BC-SENet's effectiveness in improving aviation command recognition and enhancing flight safety.
\vfill\pagebreak

\bibliographystyle{IEEEbib}
\bibliography{refs}

\end{document}